\documentclass{iau} 
\usepackage{graphicx}

\title[S316.~~Forming SSC in NGC~5253] 
{Submillimeter View of Gas and Dust in the Forming Super Star Cluster in NGC~5253}

\author[Jean L. Turner]   
{Jean L. Turner$^1$
}

\affiliation{$^1$Department of Physics and Astronomy, UCLA, \\ Los Angeles, CA 90095-1547 USA \\ email: {\tt turner@astro.ucla.edu} \\[\affilskip]
}

\pubyear{2015}
\volume{316}  
\jname{Formation, evolution, and survival of massive clusters}
\editors{C. Charbonnel \& A. Nota, eds.}
\begin{document}

\maketitle

\begin{abstract}
A giant molecular cloud has been detected surrounding
the supernebula in NGC 5253, revealing details of the formation and feedback process in
a very massive star cluster. ``Cloud D" was recently mapped in CO J=3--2 with
the Submillimeter Array. The cloud surrounds a currently forming massive cluster
of mass $\sim 10^6$~M$_\odot$, and luminosity $\sim 10^9$~L$_\odot$. Cloud D 
is hot, clearly associated with the cluster, yet kinematically relatively quiescent. The 
dust mass is $\sim$15,000~M$_\odot$, for a gas-to-dust
ratio of $\sim$50, nearly an order of magnitude lower than expected for this
low metallicity galaxy. We posit that enrichment by the cluster, leading to a stalled cluster 
wind, has created the unusual conditions in Cloud D. The absence
of current mechanical impact of the young cluster on the cloud, in spite of the presence of
thousands of O stars, may permit
future generations of stars to form near the massive cluster. 
\end{abstract}

\firstsection 
\section{Introduction}

The process by which the most massive star clusters form is not well understood. The closest
massive clusters, such as those in the Galactic Center or R136 in the Large
Magellanic Cloud, are relatively evolved and far from their initial molecular conditions. 
Clusters that are still embedded in their natal clouds can provide
more direct information 
 on the
interaction of young clusters with their natal clouds, including
 star formation efficiency and the 
precise nature of cluster feedback on the clouds. The advantage of   
submillimeter observations in the study of young clusters is that one
can simultaneously
trace spectral lines of molecular gas as well as continuum emission from ionized gas and 
dust, to give a full picture of the gas, dust, and kinematics in
young, forming clusters.

The  massive cluster in NGC~5253  may be the closest example of a embedded cluster
still  in the process of formation. Despite having little gas and no disk, this dwarf
 spheroidal galaxy has formed hundreds
of clusters in the recent past 
(\cite[Calzetti et al.~1997, 2015, Chandar et al.~2005, Harbeck et al.~2012, de Grijs et al.~2013]{Calzetti97,Chandar05,Harbeck12,deGrijs13,Calzetti15}).
This unlikely environment is the location of the ``supernebula", 
 a bright and  compact radio continuum source associated with a luminous HII region 
(\cite[Beck et al.\ 1996, Turner et al.\ 2000]{Beck96,Turner00}).
 The HII region core is  $\sim$1 pc in radius with an infrared luminosity of 
 $\sim 10^9~\rm L_\odot$ at D=3.8 Mpc, and Lyman continuum rate of $\sim 7\times 10^{52}~
 \rm s^{-1}$ (7000 O7 star equivalents) (\cite[Meier et al.\ 2002]{Meier02}). The supernebula 
 is  embedded, with $\rm A_v \sim 16$  (\cite[Turner et al. 2003,
 Alonso-Herrero et al.\ 2004]{Turner03,AH04}).
Most of the gas in NGC 5253 resides {\it outside} the galaxy, in a halo of HI filaments
\cite[(Kobulnicky \& Skillman 2008; L{\'o}pez-S{\'a}nchez et al.\ 2012)]{KS08,LS12}
and in a molecular streamer of $\sim$200 pc extent detected in 
CO (\cite[Turner et al.\ 1997]{Turner97}). The streamer is infalling along the minor axis
of the galaxy.
No CO(1--0) and only weak CO(2--1) had been detected until recently
at the location of the embedded SSC/supernebula 
(\cite[Turner et al.\ 1997; Meier et al.\ 2002; Miura et al.\ 2015]{Turner97,Meier02,Miura15}).

\begin{figure}[b]
\begin{center}
 \includegraphics[width=4in]{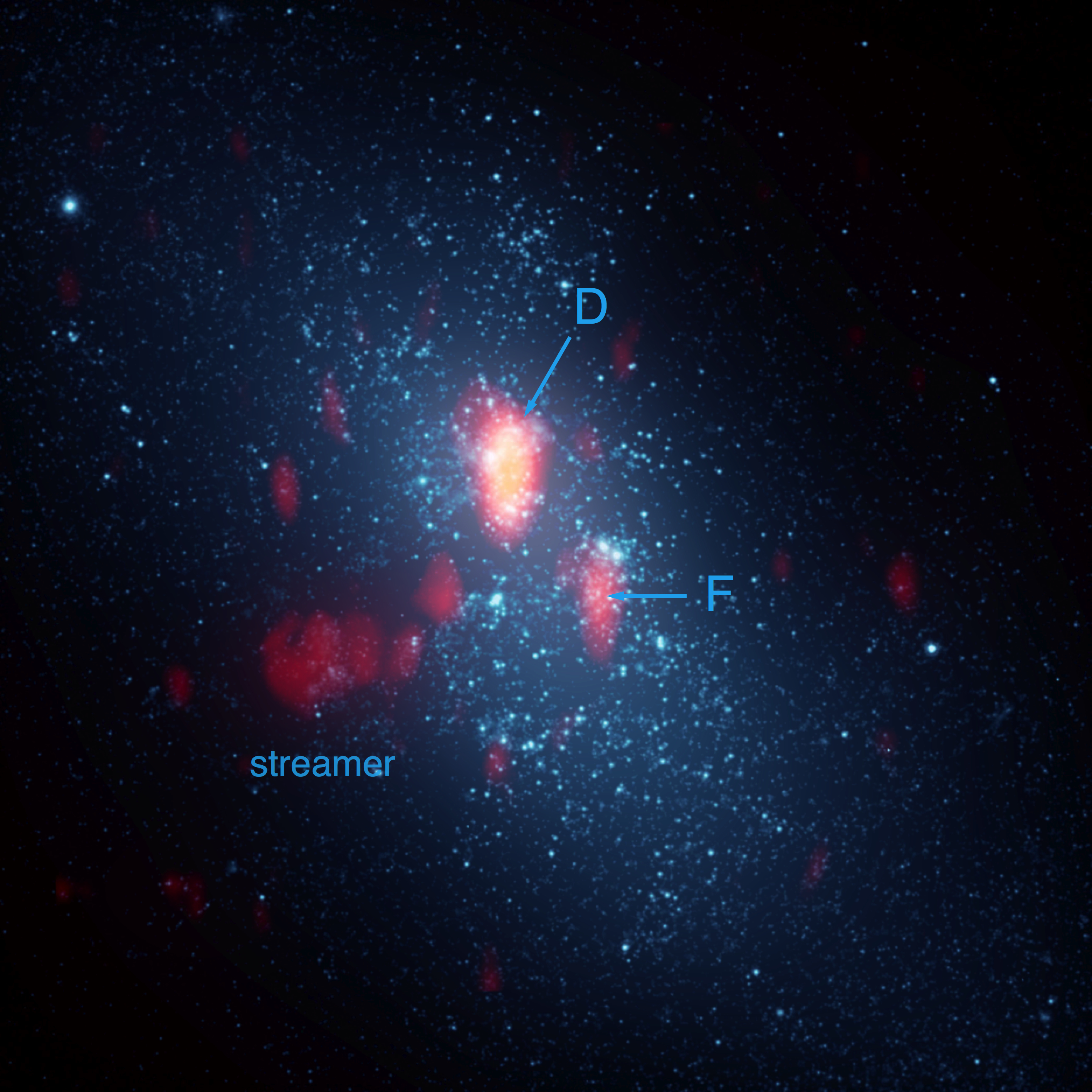} 
 \caption{Molecular gas in NGC 5253. CO(3-2) emission overlaid on an
 HST image. The CO image is from the SMA, beam is 4$^{\prime\prime}$ x 2
$^{\prime\prime}$. ``Cloud D" is the brightest CO source near the center; also
detected are another cloud, ``F" and the
``streamer", which is the extended filament to the east. Turner et al.\ 2015.}
   \label{fig1}
\end{center}
\end{figure}

Submillimeter Array observations of the J=3--2 transition of CO and dust continuum in
NGC 5253 have revealed the giant molecular cloud (GMC)
associated with the supernebula and its embedded cluster.
``Cloud D" is shown in 
Figure\,\ref{fig1}  \cite[(Turner et al.\ 2015)]{Turner15}. 
Cloud D is unusually bright in CO(3--2), 2.6 times brighter than in CO(2--1).
The emission is optically thin and the inferred gas temperature is $>$200~K, with 
densities of  $\sim 5\times 10^4~\rm cm^{-3}$.  Galactic GMCs, even those actively forming
stars, more typically 
have temperatures of $\sim$10-15~K on these scales. From its temperature
alone, it is clear that Cloud D is  closely
associated with the young massive cluster.

To determine a star formation 
efficiency for the cluster we need a gas mass, which is difficult 
to determine for this strange cloud. From the CO(3--2)/CO(1--0) ratio it is inferred that the
emission is optically thin;  by contrast, Galactic GMCs are optically thick in
CO.  Thus we cannot use the standard Galactic conversion factor, 
$X_{CO}$, which applies to optically thick and   cooler 
Galactic clouds in the CO(1--0) transition. Cloud D is as yet undetected in CO(1--0). 
For optically thin emission, instead one counts emitted photons
in the line and from that computes the total
number of CO molecules from the partition function. However, to get to the total molecular
mass requires  a relative abundance of CO to H$_2$, and this abundance
 depends on the chemistry. The
chemistry is uncertain in Cloud D because of the unusually high temperature of the cloud,
the high radiation fields, and the low metallicity of NGC~5253, $Z\sim 0.25Z_\odot$ 
(\cite[Kobulnicky et al.\ 1997, L{\'o}pez-S{\'a}nchez et al.\ 2007]{Kobulnicky97,LS07}).
Using the Galactic value of $\rm [CO]/[H_2]=8.5\times 10^{-5}$ we  infer a total
H$_2$ mass of $M_{H_2}=5\times10^4~\rm M_\odot$ \cite[(Turner et al.\ 2015)]{Turner15}; 
however, given the unusual properties of this cloud we have no confidence that the 
  Galactic CO 
abundance ratio obtains here.

A  less slippery number for the gas mass can be determined from subtracting
the  stellar mass from the  dynamical
mass of the cloud. The
uncertainties are simpler and better constrained than the previous methods; it gives
a total gas mass rather than an H$_2$ mass, which is more appropriate
for star formation efficiency. 
The mass of stars  can be obtained from the Lyman continuum
rate (\cite[Meier et al.\ 2002]{Meier02}) and the 3.5 Myr age based on Br $\gamma$ 
equivalent widths (\cite[Alonso-Herrero et al.\ 2004]{AH04}) and 
Wolf-Rayet signatures in abundances (\cite[Walsh \& Roy 1989, Kobulnicky et al.\ 1997,
L{\'o}pez-S{\'a}nchez et al.\ 2007]{WR89, Kobulnicky97, LS07}). 
\cite[Calzetti et al.\ (2015)]{Calzetti15} argue that dust emission affects the Br $\gamma$
EW and that the age of this cluster is closer to 1 Myr, and further divide the stars among
two clusters, both within Cloud D. However, this young age would not explain
the Wolf-Rayet signatures and local nitrogen enhancement in the region, so we favor
an older age. For a 3.5 Myr age, and assuming continuous star formation, the stellar
mass based on {\small STARBURST99} models
is $M_{stars}=1.1_{-0.2}^{+0.7} \times 10^6~\rm M_{\odot}$. The CO linewidth is 
$\sigma = 9.2 \pm 0.6~\rm km\,s^{-1}$; this will give an upper limit to the gravitational
mass, which is $M_{vir} = 1.8_{-0.7}^{+0.2}\times 10^6~\rm M_{\odot}$. The final
gas mass is then $M_{vir}-M_{stars} =7\pm4\times 10^5~\rm M_{\odot}$, giving
a star formation efficiency $M_{stars}/(M_{stars}+M_{gas} )\sim 60$\%
\cite[(Turner et al.\ 2015)]{Turner15}. This very
high, record efficiency, not so different from the 75$\pm$50\% inferred for Cloud D from CO(2--1)
using $X_{CO}$ by \cite[Meier et al. (2002)]{Meier02}, bodes well for the cluster's survival.

\section{Cloud D: A Very Dusty Cloud}
An alternative path to the gas mass is through dust, but the dust mass of Cloud D
is also strange and interesting. About half of the 72 mJy of 870$\mu$m continuum
emission, 34$\pm$14 mJy, is from dust; the rest is free-free emission from
the ionized gas. Using the dust opacity
of the Large Magellanic Cloud (\cite[Galliano et al.\ 2011]{Galliano11}), and a 
dust temperature of 46 K (\cite[Thronson \& Telesco 1986]{TT86}) we obtain
$M_{dust}=1.5\pm 0.2 \times 10^4~\rm M_{\odot}.$ If we scale the gas-to-dust ratio, $GTD$,
of NGC~5253 to match its $Z\sim 0.25Z_\odot$  metallicity, a common and reasonable assumption,
 we would estimate that $GTD\sim 500-700$, similar to values in the Magellanic Clouds, 
and this would give a dust-predicted gas mass of 
$M_{gas}\sim 10^7~\rm M_{\odot}.$ This is more than 
five times higher than the gas mass could possibly be, based on the dynamical
mass of Cloud D. From the inferred gas mass we infer that
instead $GTD\sim 50$ in NGC~5253, a factor of three less than the Galactic $GTD$ value 
\cite[(Turner et al.\ 2015)]{Turner15}. 

The reason that  NGC 5253 has ten times more dust than expected may be enrichment
by the cluster stars, metals that are as yet trapped within the natal cloud. {\small
STARBURST99} models for clusters of this luminosity that are  more than 3 Myr of age
can produce 30,000 $\rm M_\odot$ of metals, twice the dust mass that is seen. So
it is not surprising, it is even expected, that enrichment is taking place. 

The dust enrichment has potential implications for the evolution of the cluster and its
cloud, and the nature of feedback. Cloud D is surprisingly quiescent; its linewidth
of $\sim$9~$\rm km\, s^{-1}$ is not far from the value expected for a cloud of its 
25 pc size based on the Galactic size-linewidth relation. Yet its high temperature and
precise positional and kinematic coincidence with the massive cluster indicates a 
close association of the two. Where is the feedback from the 7000 O stars? 

\cite[Silich et al.\ (2004) and Tenorio-Tagle et al.\ (2007)]{Silich04,TT07} suggest that
enrichment by cluster stars creates 
a dusty environment that can enhance gas cooling and stall a cluster wind. Based on 
numerical modeling of the gas dynamics they find that
for a cluster of the luminosity exciting the supernebula in NGC~5253, an 
enrichment of 1.5$Z_\odot$ is sufficient to stall a  wind, trapping the enriched
products of stellar evolution within the cloud. From the mass of dust it would
appear that the enrichment of the gas in Cloud D is roughly twice this amount. The
stalled wind model
could explain both the quiescence of Cloud D and its dustiness. Given the relative 
quiescence of this cloud, so similar to Galactic GMCs, it would seem that 
conditions are ripe for the formation of even more stars at this location, given an 
influx of gas. This is a very interesting suggestion 
\cite[(Silich et al.\ 2004; Tenorio-Tagle et al.\ 2007)]{Silich04,TT07} in light of
the fact that multiple generations are observed to be present in globular clusters
(see also R. W{\"u}nsch this volume). 

\section{Conclusions}

Submillimeter Array maps of CO(3--2) and 870$\mu$m dust continuum emission from
Cloud D associated with the young,
embedded massive cluster exciting the supernebula in NGC 5253 allow the determination
of dynamical mass, gas mass, and dust mass for this cloud, and allow us to define
the interaction of the cluster
and its surroundings.

1) The molecular gas in Cloud D is  hot; the  CO(3--2)/CO(2--1) line ratio suggests that the
temperature is unusually warm, $>$ 200~K. The cloud is clearly heated by the cluster.

2) The H$_2$ mass obtained from the line intensity of the optically thin CO emission is low, 
particularly compared to the dust mass. This is
probably due to the [CO]/[H$_2$] abundance ratio, which is highly uncertain in this unusually
warm and highly irradiated cloud. Instead we obtain a 
 gas
mass by subtracting the stellar mass, based on the Lyman continuum rate and 
{\small STARBURST99} models and a 3.5 Myr age, 
from the dynamical mass based on the CO size and linewidth
of Cloud D. The gas mass of $7\times 10^5~\rm (D/3.8\,Mpc)^2$ gives
 a record high star formation efficiency of 60\%. 

3) The dust mass is  high for the metallicity of the galaxy: gas-to-dust radio is 50:1, sub-solar, 
 much lower than the 500-700 value expected based on the low metallicity of NGC 5253.
We posit that the dust mass is almost entirely produced by the stars of the cluster; 
{\small STARBURST99} models predict twice this yield in metals for a cluster of this
luminosity and 3.5 Myr age, so this model is entirely plausible.

4) The numerical models of \cite[Silich et al.\ (2004)]{Silich04} predict that a cluster
wind will catastrophically cool and stall if the gas is enriched to 1.5 $Z_\odot$
for the supernebula cluster in NGC 5253. This condition 
appears to be met. This could explain why, even in the presence of 7000 O stars
and a high cloud temperature, the width of the CO line, 9.2 km/s, roughly that expected for a 
quiescent cloud of its size based on the Galactic size-linewidth relation. 
Cluster feedback of the mechanical variety appears to be 
inhibited in this cluster, which could allow further accretion and future generations
of stars to form in this cluster.

\acknowledgements
JLT would like to acknowledge S. Michelle Consiglio for helpful discussions, and
the support of NSF Grant AST-1515570.

\begin{discussion}

\discuss{Boesgaard}{How did you get the metallicity?}

\discuss{Turner}{The overall metallicity of NGC~5253, $Z\sim0.25 Z_\odot$, 
 is from observations of the surrounding
HII regions. Optical emission lines cannot be detected from the supernebula
directly, which
is embedded in Cloud D. The extinction is about 16 magnitudes in the visual.
The gas-to-dust ratio is directly computed from the gas mass of Cloud D and 
observed dust mass.}

\end{discussion}

\end{document}